\documentclass[11pt,aps,showpacs,showkeys]{revtex4}
\usepackage{amsmath,amsthm,amssymb,epsfig,alltt}

\begin{document}
	
	\def\a{\alpha}
	\def\b{\beta}
	\def\d{{\delta}}
	\def\l{\lambda}
	\def\e{\epsilon}
	\def\p{\partial}
	\def\m{\mu}
	\def\n{\nu}
	\def\t{\tau}
	\def\th{\theta}
	\def\s{\sigma}
	\def\g{\gamma}
	\def\o{\omega}
	\def\r{\rho}
	\def\z{\zeta}
	\def\D{\Delta}
	\def\half{\frac{1}{2}}
	\def\hatt{{\hat t}}
	\def\hatx{{\hat x}}
	\def\hatp{{\hat p}}
	\def\hatX{{\hat X}}
	\def\hatY{{\hat Y}}
	\def\hatP{{\hat P}}
	\def\haty{{\hat y}}
	\def\whatX{{\widehat{X}}}
	\def\whata{{\widehat{\alpha}}}
	\def\whatb{{\widehat{\beta}}}
	\def\whatV{{\widehat{V}}}
	\def\hatth{{\hat \theta}}
	\def\hatta{{\hat \tau}}
	\def\hatrh{{\hat \rho}}
	\def\hatva{{\hat \varphi}}
	\def\barx{{\bar x}}
	\def\bary{{\bar y}}
	\def\barz{{\bar z}}
	\def\baro{{\bar \omega}}
	\def\barpsi{{\bar \psi}}
	\def\sp{\sigma^\prime}
	\def\nn{\nonumber}
	\def\cb{{\cal B}}
	\def\2pap{2\pi\alpha^\prime}
	\def\pap{\pi\alpha^\prime}
	\def\wideA{\widehat{A}}
	\def\wideF{\widehat{F}}
	\def\beq{\begin{eqnarray}}
		\def\eeq{\end{eqnarray}}
	\def\4pap{4\pi\a^\prime}
	\def\op{\omega^\prime}
	\def\xp{{x^\prime}}
	\def\sp{{\s^\prime}}
	\def\ap{\a^\prime}
	\def\tp{{\t^\prime}}
	\def\zp{{z^\prime}}
	\def\rp{{\rho^\prime}}
	\def\spp{\s^{\prime\prime}}
	\def\xpp{x^{\prime\prime}}
	\def\xppp{x^{\prime\prime\prime}}
	\def\barxp{{\bar x}^\prime}
	\def\barzp{{\bar z}^\prime}
	\def\barxpp{{\bar x}^{\prime\prime}}
	\def\barxppp{{\bar x}^{\prime\prime\prime}}
	\def\zetap{{\zeta^\prime}}
	\def\barchi{{\bar \chi}}
	\def\baro{{\bar \omega}}
	\def\bpsi{{\bar \psi}}
	\def\barg{{\bar g}}
	\def\barz{{\bar z}}
	\def\bareta{{\bar \eta}}
	\def\ta{{\tilde \a}}
	\def\tb{{\tilde \b}}
	\def\tc{{\tilde c}}
	\def\tz{{\tilde z}}
	\def\tJ{{\tilde J}}
	\def\tpsi{\tilde{\psi}}
	\def\tal{{\tilde \alpha}}
	\def\tbe{{\tilde \beta}}
	\def\tga{{\tilde \gamma}}
	\def\tchi{{\tilde{\chi}}}
	\def\barth{{\bar \theta}}
	\def\bareta{{\bar \eta}}
	\def\barom{{\bar \omega}}
	\def\bole{{\boldsymbol \epsilon}}
	\def\bolth{{\boldsymbol \theta}}
	\def\bomega{{\boldsymbol \omega}}
	\def\bolmu{{\boldsymbol \mu}}
	\def\bolal{{\boldsymbol \alpha}}
	\def\bolbe{{\boldsymbol \beta}}
	\def\bolL{{\boldsymbol  L}}
	\def\bolX{{\boldsymbol X}}
	\def\bbk{{\boldsymbol k}}
	\def\boln{{\boldsymbol n}}
	\def\bols{{\boldsymbol s}}
	\def\bolS{{\boldsymbol S}}
	\def\bola{{\boldsymbol a}}
	\def\bolA{{\boldsymbol A}}
	\def\boleta{{\boldsymbol \eta}}
	\def\bolchi{{\boldsymbol \chi}}
	\def\bolJ{{\boldsymbol J}}
	\def\tr{{\rm tr}}
	\def\bbP{{\mathbb P}}
	\def\bbp{{\boldsymbol p}}
	\def\mathP{{\mathbb P}}

	\setcounter{page}{1}
	\title[]{BRST Ghost-Vertex Operator in Witten's Cubic Open String Field Theory on Multiple $Dp$-branes}

	\author{Taejin Lee}
	\email{taejin@kangwon.ac.kr}
	\affiliation{Department of Physics, Kangwon National University, Chuncheon 24341 Korea}
	
	\date{\today }

	\begin{abstract}

		The Becchi--Rouet--Stora--Tyutin (BRST) ghost field is a key element in constructing Witten's cubic open string field theory. However, to date, the ghost sector of the string field theory has not received a great deal of attention. 
		In this study, we address the BRST ghost on multiple $Dp$-branes, which carries non-Abelian indices and couples to a non-Ablelian gauge field. We found that the massless components of the BRST ghost field can play the role of the Faddeev--Popov ghost in the non-Alelian gauge field, such that the string field theory maintains the local non-Abelian gauge invariance.

	\end{abstract}

	
	
	
	\maketitle
	
	\setcounter{footnote}{0}

	\newpage
	\tableofcontents
	\newpage

	\section{Introduction}
	
	In a recent study \cite{Lee2022cubicm}, we extended the Witten open string field theory \cite{Witten1986,Witten92p} on a single $D25$-brane to a cubic open string field theory on multiple $Dp$-branes, $p= -1, 0, \cdots, 25$. On multiple $Dp$-branes, both the string field and gauge parameters carry non-Abelian group indices. We expect that the Faddeev--Popov ghost \cite{Faddeev} structure originates from the low-energy sector of the BRST ghost field. Siegel \cite{Siegel1985aPLB,Siegel1985bPLB} pointed out that the massless component of the BRST ghost field could be the Faddeev--Popov ghost of gauge theory. However, his discussion was limited to the $U(1)$ gauge field \cite{Banks1986}, which is the low-energy sector of open string field theory on a single $D25$-brane. 
	
	In non-Abelian gauge theory, which describes the low-energy sector of an open string on multiple $Dp$-branes, the Faddeev--Popov ghost field interacts with the non-Abelian gauge field. Therefore, it is necessary to examine cubic string coupling to confirm that the Faddeev--Popov ghost structure is consistent with the non-Abelian gauge symmetry of the low-energy sector of strings on multiple $Dp$-branes. For this purpose, 
	we construct the BRST ghost-vertex operator for Witten's cubic open string field theory on multiple $Dp$-branes. Because the BRST ghost fields transform nontrivially under conformal transformation, it is not easy to find a propagator on the string world sheet to evaluate the Polyakov string path integral, which leads us to the cubic string vertex in the 
	ghost sector. 
	
	In this study, we construct the three-string vertex operator in the ghost sector 
	explicitly for Witten's cubic open string field theory for multiple $Dp$-branes. In the next section, we define Witten's cubic open string field theory on multiple $Dp$-branes with the 
	overlapping functions of the string and BRST ghost coordinates. 
	We could directly convert the overlapping functions into their Fock space representations
	\cite{kaku74a,kaku74b,Ohta1986,Grossjevicki87a,Grossjevicki87b,Samuel86} to obtain the vertex operators. However, this procedure requires inverting infinite-dimensional matrices that could possibly not be uniquely defined. Moreover, the obtained vertex operators could possibly not correctly reproduce the scattering amplitudes represented by the Polyakov string path integrals. These problems were identified earlier by 
	Cremmer and Gervais \cite{cremmer1975}.
	To avoid these problems, we apply a different strategy, called the Mandelstam procedure \cite{Mandelstam1973,Mandelstam1974,Hata86,Hata1986}, advocated for the light-cone string field theory by Mandelstam, and map the string world sheet defined by overlapping functions onto the upper or lower half complex planes (in the case of a closed string, full complex plane), where the propagators of the coordinate fields take simple forms. 
	Conformal mapping from the string world sheet onto the complex plane is called Schwarz--Christoffel mapping. 
	
	Using Schwarz--Christoffel mapping and simple propagators (Green's function) on the complex plane, we can identify the Fourier components of Green's function, called the
	Neumann functions, on the string world sheet, which are typically given as contour integrals. We provide a precise definition of the Neumann functions for ghost coordinates and explicitly calculate specific examples of these. As we will see, these explicit values of the Neumann functions of the ghost coordinates are crucial for proving the massless components of the BRST ghost fields.

	\section{Witten's Cubic Open String Field Theory with BRST Ghost Fields on Multiple $Dp$-branes}

	Witten's cubic open string field theory on multiple $Dp$-branes is described by the following action:
	\beq
	S = \int {\rm tr} \left(\Psi*Q\Psi + \frac{2}{3} \Psi * \Psi * \Psi \right),
	\eeq 
	where $Q$ is the BRST operator. 
	Here, the star product $*$ with the BRST ghost coordinates is defined as 
	\beq
	\int \overbrace{\Psi*\Psi* \cdots *\Psi}^M &=& \int \prod^M_{r=1} D[X^{(r)}(\s)]\prod_{r=1}^M \prod_{0\le \s \le\pi/2} \d\left[X^{(r)}(\s)-X^{(r-1)}(\pi-\s)\right] \nn\\
	&& \int \prod^N_{r=1} D[c^{(r)}(\s)]\prod_{r=1}^M \prod_{0\le \s \le\pi/2} \d\left[c^{(r)}(\s)+c^{(r-1)}(\pi-\s)\right] \nn\\
	&& {\rm tr} \left(\prod_{r=1}^M\Psi[X^{(r)}(\s), c^{(r)}(\s)]\right).
	\eeq 
	This action is invariant under the BRST gauge transformation 
	\beq
	\d \Psi = Q*\e - \e * \Psi + \Psi * \e.
	\eeq 
	On multiple $Dp$-branes, both string field $\Psi$ and gauge parameter field $\e$ can carry $U(N)$ group indices.
	
	\subsection{BRST ghost coordinates of open string field theory}
	
	The ghost part of the Polyakov string path integral and action are given by 
	\beq
	{\cal A}_{\rm gh} &=& \int D[b,c] D[\tilde b, \tilde c] \exp \left( S_{\rm ghost} \right), \nn\\
	S_{\rm ghost} &=& \frac{1}{2\pi} \int dz d \bar z\left( b \frac{\p}{\p \bar z} c + \tilde b \frac{\p}{\p z} \tilde c\right).
	\eeq 
	The ghost coordinates $b$ and $c$ are Grassmann-odd fields on the two-dimensional space of conformal dimensions $(2,0)$ and $(-1,0)$.
	To evaluate the Polyakov string path integral and construct the vertex operators,
	we must know the propagators of the ghost fields on the string world sheet. 
	However, it is difficult to construct the propagators of the ghost fields of open strings 
	directly on the string world sheet. We know that on the complex plane (closed string), the holomorphic and anti-holomorphic parts of the propagator are given as 
	\beq
	\langle b(z) c(z^\prime) \rangle = \frac{1}{z - \zp},   ~~~\langle \tilde b( \bar z) \tilde c( \bar z^\prime) \rangle = \frac{1}{\bar z - \bar z^\prime}.
	\eeq 
	Thus, we must identify how the BRST ghost fields of an open string transform under conformal transformation. 
	
	To understand the conformal transformation of open string BRST ghost fields, it is 
	convenient to consider an open string as a folded closed string \cite{Lee1988Ann}. The folding condition is 
	as follows (on a cylindrical surface): 
	\beq
	b(\t,\s) &=& \sum_n b_n e^{-n(\t+i\s)} = \tilde b(\t,-\s) = \sum_n \tilde b_n e^{-n(\t+i\s)}, \nn\\
	c(\t,\s) &=& \sum_n c_n e^{-n(\t+i\s)} = \tilde c(\t,-\s) = \sum_n \tilde c_n e^{-n(\t+i\s)},
	\eeq 
	which are equivalent to 
	\beq
	b_n = \tilde b_n, ~~~c_n = \tilde c_n.
	\eeq 
	Considering these folding conditions, we can define the 
	ghost fields of open strings as follows: 
	\beq
	b(\t,\s)_{\rm open}  &=& \half \left\{ b(\t, \s) + \tilde b(\t,\s) \right\} = \sum_n b_n e^{-n\t} \cos n\s, \nn\\
	c(\t,\s)_{\rm open}  &=& \half \left\{ c(\t, \s) + \tilde c(\t,\s) \right\} = \sum_n c_n e^{-n\t} \cos n\s.
	\eeq 
	
	\subsection{Propagators of $b$-$c$ ghost fields of open string} 
	
	From the folding condition, it follows that Green's function (propagator) of the BRST ghost fields of the open string can be expressed as 
	\beq
	\langle b(z) c(\zp) \rangle_{\rm open}  = \frac{1}{4} \left\{\langle b(z) c(\zp) \rangle
	+ \langle b(z) \tilde c(\barzp) \rangle + \langle \tilde b(\barz) c(\zp) \rangle + 
	\langle \tilde b(\barz) c(\barzp) \rangle
	\right\}.
    \eeq 
With the help of the folding procedure, we easily find the free Green's function of the $b$–$c$ ghost fields of an open string on the upper half (or lower half) complex plane satisfying the Neumann boundary condition as 
\beq
\langle b(z) c(\zp) \rangle_{\rm open} =  \frac{1}{4} \left\{\frac{1}{z-\zp} + \frac{1}{\barz-\zp} + \frac{1}{z-\barzp} + \frac{1}{\barz-\barzp}
\right\}.
\eeq

If the folding condition is imposed, the string world sheet of the closed string, which is a cylindrical surface for the 
free string, is folded into a strip, which corresponds to the string world sheet of an open string. 
The conformal mapping from the complex plane onto the cylindrical surface (the world sheet of a free closed string) is given by 
\beq
z = f(\z) = e^\zeta, ~~~ \bar z = f(\bar \z) = e^{\bar \z}. 
\eeq 
Under this conformal mapping, the $b$-$c$ ghost fields transform as 
\beq
b(z) \rightarrow b(\z) = b(z) \left(\frac{\p f}{\p \z}\right)^2= b(z) z^2, ~~
c(z) \rightarrow c(\z) = c(z) \left(\frac{\p f}{\p \z}\right)^{-1}= \frac{c(z)}{z}. 
\eeq 
It follows from the conformal transformation of the $b$-$c$ ghost coordinates and 
folding construction that the free propagator of the $b$–$c$ ghost fields of an open string on a strip must be given as 
\beq
\langle b(z) c(\zp) \rangle_{\rm open} 
&=& \frac{1}{4} \left\{\frac{z^2 (\zp)^{-1}}{z-\zp} + \frac{(\barz)^2 (\zp)^{-1}}{\barz-\zp} + \frac{z^2 (\barzp)^{-1}}{z-\barzp} + \frac{(\barz)^2(\barzp)^{-1}}{\barz-\barzp}
\right\}.
\eeq

Green's function of the $b$–$c$ ghost fields of an open string on a general string world sheet can be obtained in a similar manner if the conformal mapping from the 
complex plane onto the string world sheet (or conformal mapping from the string world sheet onto the complex plane)
of the interacting open strings can be found. If the conformal mapping from each patch (local coordinates) $z_r(\z_r)$ to the complex plane is found, Green's function of the $b$–$c$ ghost fields is constructed as 
\beq \label{Grbrst}
G(\z_r, \zp_s) &=& \langle b(\z_r) c(\zetap_s) \rangle_{\rm world-sheet} \nn\\
&=&
\frac{1}{4} \left\{\left(\frac{\p z_r}{\p \zeta_r}\right)^{2} \left(\frac{\p \zp_s}{\p \zeta_s^\prime}\right)^{-1} 
\frac{1}{z_r-\zp_s} + \left(\frac{\p z_r}{\p \zeta_r}\right)^{2} \left(\frac{\p \barzp_s}{\p \bar \zeta^\prime_s}\right)^{-1} \frac{1}{z_r-\barzp_s} + C.C. \right\}.
\eeq 
Details of the construction of Green's function and the calculations of the Neumann functions of the $b$–$c$ ghost fields are given in the Appendix.

\section{Construction of Vertex Operators for BRST ghost}

\begin{figure}[htbp]
	\begin {center}
	\epsfxsize=0.7\hsize
	
	\epsfbox{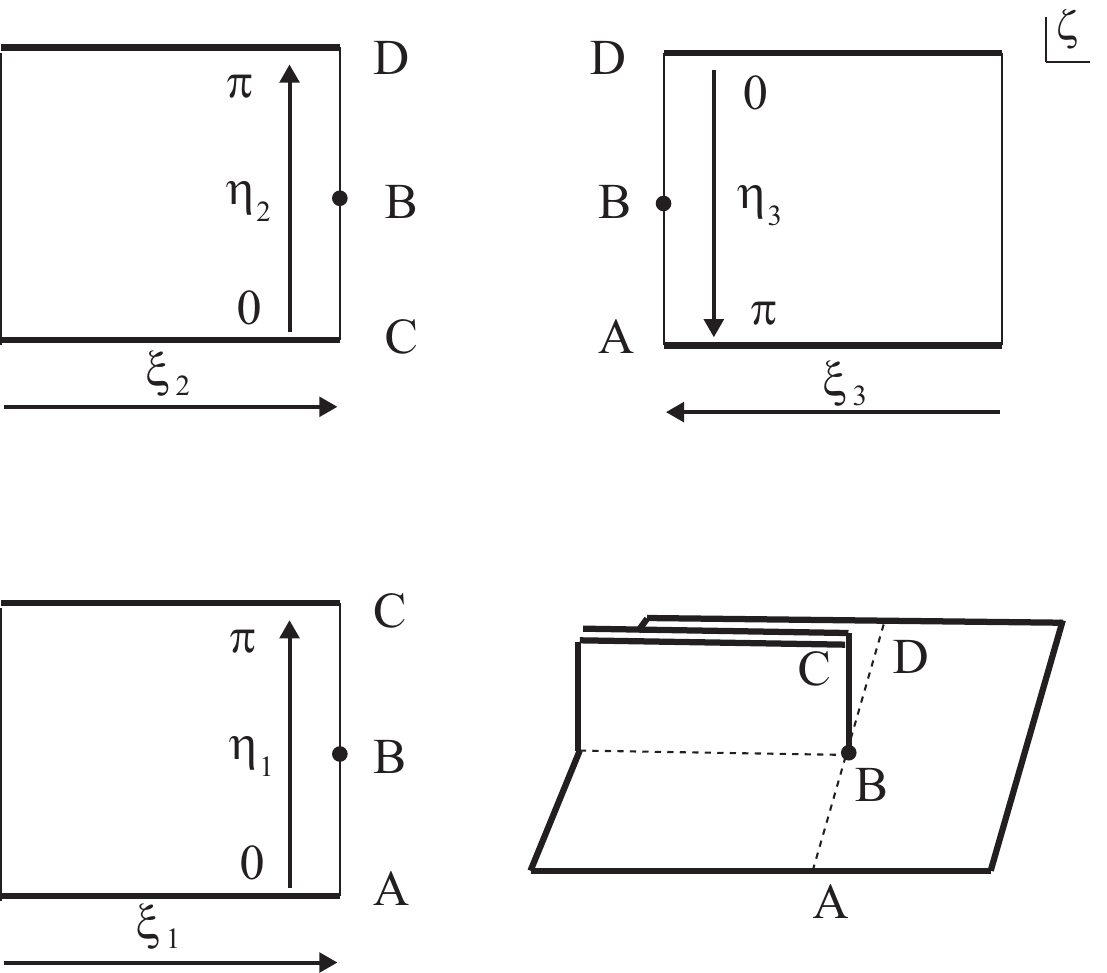}
	\end {center}
	\caption {\label{3patches1} World-sheet of three-open-string interaction.}
\end{figure}

Assuming that we found Green's function of the ghost fields on the string world sheet, in this section, we construct the vertex operator for the 
BRST ghost fields. The world sheet of the three-open-string interaction is displayed in Fig. \ref{3patches1}. \cite{TLeeJKPS2017}

We can use Green's function for the ghost field on the string world sheet as 
\beq \label{Green15}
G(\eta_r, \xi_r; \eta^\prime_s, \xi_s^\prime) &=& \d_{rs} \sum_{n \ge -1} \cos n \eta_r \cos n \eta^\prime_s  + \sum_{n, m \ge -1} \bar G^{rs}_{nm} e^{|n|\xi_r + |m|\xi_s} \cos n \eta_r \cos m \eta^\prime_s.
\eeq 
Here, $\z_r = \xi_r + i \eta_r$ and $ - \infty < \xi_r \le 0$, $0 \le \eta_r \le \pi$, $ r = 1, 2, 3$. The first part of Green's function
is a result of the free propagator. 

With the boundary values of the ghost fields fixed by
$\{b^{(r)}, c^{(r)}, r= 1, 2, 3   \}$, we can evaluate the Polyakov string path integral, which defines the scattering amplitude as
\beq
\exp\left(F_{\rm ghost}\right) &=& \int D[b,c]  \exp \left\{
S_{\rm ghost} + \sum_r \int_{\p M^{(r)}} \left(b c^{(r)} + b^{(r)} c \right)\right\}, 
\eeq 
where $\p M^{(r)}$ and $r=1,2,3$ are the temporal boundaries of the string world sheet. 
Using Green's function, Eq. (\ref{Green15}), we can rewrite $F_{\rm ghost}$ as 
\beq \label{Fghost}
F_{\rm ghost} 
&=& \frac{1}{\pi^2} \sum_{r, s}  \int_0^\pi d \eta_r \int_0^\pi d\eta_s 
\left\{b^{(r)}_0 + \sum_{n=1} b^{(r)}_n \cos n \eta_r \right\}  G(\eta_r,\t_r; \eta^\prime_s, \t^\prime_s)\nn\\
&&
\left\{c^{(s)}_0 + \sum_{m=1} c^{(s)}_n \cos m \eta^\prime_s \right\}\nn\\
&=& \sum_r \left\{b^{(r)}_0 c^{(r)}_0 + \half  
\left( b_1^{(r)} + b_{-1}^{(r)} \right)  \left( c_1^{(r)} + c_{-1}^{(r)} \right) + \frac{1}{4}\sum_{n \ge 2} \left( b_n^{(r)} + b_{-n}^{(r)} \right) \left( c_n^{(r)} + c_{-n}^{(r)} \right) 
\right\} \nn\\
&& + \sum_{r,s} \Biggl\{\frac{1}{2}\bar G^{rs}_{00} b^{(r)}_0 c^{(s)}_0 + \frac{1}{2\sqrt{2}}  \bar G^{rs}_{(-1)0}e^{-\xi_r} \left( b_1^{(r)} + b_{-1}^{(r)} \right) c^{(s)}_0 + \frac{1}{2\sqrt{2}} \bar G^{rs}_{0(-1)}e^{-\xi_s} b^{(r)}_0\left( c_1^{(r)} + c_{-1}^{(r)} \right) \nn\\
&&+ \frac{1}{\sqrt{2}} \sum_{n\ge 1} \bar G^{rs}_{n0} e^{n \xi_r} \left( b_n^{(r)} + b_{-n}^{(r)} \right) c^{(s)}_0 
+ \frac{1}{\sqrt{2}} \sum_{m\ge 1} \bar G^{rs}_{0m} e^{m \xi_s} b^{(r)}_0 \left( c_m^{(s)} + c_{-m}^{(s)} \right) \nn\\
&& + \frac{1}{4} \bar G^{rs}_{(-1)(-1)} e^{-\xi_r -\xi_s} \left( b_1^{(r)} + b_{-1}^{(r)} \right)  \left( c_1^{(s)} + c_{-1}^{(s)} \right) 
\nn\\
&& + \frac{1}{2} \sum_{m \ge 1} \bar G^{rs}_{(-1)m} e^{-\xi_r +m\xi_s} \left( b_1^{(r)} + b_{-1}^{(r)} \right)  \left( c_m^{(s)} + c_{-m}^{(s)} \right) \nn\\
&&+ \frac{1}{2} \sum_{n \ge 1} \bar G^{rs}_{n(-1)} e^{n\xi_r -\xi_s} \left( b_n^{(r)} + b_{-n}^{(r)} \right)  \left( c_1^{(s)} + c_{-1}^{(s)} \right) \nn\\
&& +  \sum_{n,m \ge 1} \bar G^{rs}_{nm} e^{n\xi_r +m\xi_s} \left( b_n^{(r)} + b_{-n}^{(r)} \right)  \left( c_m^{(s)} + c_{-m}^{(s)} \right)
\Biggr\}.
\eeq

Now, we rewrite this expression of $F_{\rm ghost} $ into an operatorial form.
It is useful to consider a set of anticommuting operators $\{b_{1}, b_{-1}, c_1, c_{-1}\}$ satisfying 
\beq 
\{b_{1}, b_{-1}\} &=& 0, ~~\{b_1, c_{1} \} = 0, ,~~\{b_1, c_{-1} \} = 1, \nn\\
\{b_{-1}, c_{1}\} &=& 1, ~~\{c_{-1}, b_{-1}\} =0, \{c_1, c_{-1} \} = 0. 
\eeq 
We can construct a coherent state for these anticommuting operators with a set of eigenvalues $\th$ and $\chi$ as follows:
\beq \label{coh1}
\left(b_{1} + b_{-1} \right) \vert \th, \chi \rangle &=& \chi \vert \th, \chi \rangle,~~ \left(c_1 + c_{-1}\right)\vert \th, \chi \rangle = \th \vert \th, \chi \rangle, ~~ \nn\\
\vert \th, \chi\rangle 
&=&  \exp \left\{ \half \th \chi + b_{-1} c_{-1}+ \th b_{-1} + \chi c_{-1} \right\} \vert 0 \rangle.
\eeq 
By comparing the coherent states in Eq. (\ref{coh1}) with $F_{\rm ghost}$, the first line of Eq. (\ref{Fghost}) must correspond to the normalization factor of the corresponding coherent state.
\beq
\prod_r \langle b^{(r)}, c^{(r)} \vert 0 \rangle &=& \exp \Biggl\{  \sum_r \Biggl(b^{(r)}_0 c^{(r)}_0 + \half  
\left( b_1^{(r)} + b_{-1}^{(r)} \right)  \left( c_1^{(r)} + c_{-1}^{(r)} \right) \nn\\
&&
+ \frac{1}{4}\sum_{n \ge 2} 
\left( b_n^{(r)} + b_{-n}^{(r)} \right) \left( c_n^{(r)} + c_{-n}^{(r)} \right) 
\Biggr) \Biggr\}
\eeq

Using the coherent state $\prod_r \vert b^{(r)}, c^{(r)} \rangle$, we can rewrite the Polyakov string path integral in the ghost sector in an operatorial form:
\beq
\exp\left(F_{\rm ghost}\right) &=& \prod_r \langle b^{(r)}, c^{(r)} \vert \exp\left(\hat F^\prime_{\rm ghost}\right) \vert 0 \rangle, \nn\\
\hat F^\prime_{\rm ghost} &=& \sum_{r,s} \Biggl\{\frac{1}{2}\bar G^{rs}_{00} \hat b^{(r)}_0 \hat c^{(s)}_0 + \frac{1}{2\sqrt{2}} \bar G^{rs}_{(-1)0}e^{\xi_r} \left( \hat b_1^{(r)} + \hat b_{-1}^{(r)} \right) \hat c^{(s)}_0 + \frac{1}{2\sqrt{2}} \bar G^{rs}_{0(-1)}e^{\xi_s} \hat b^{(r)}_0\left( \hat c_1^{(r)} + \hat c_{-1}^{(r)} \right) \nn\\
&&+ \frac{1}{\sqrt{2}} \sum_{n\ge 1} \bar G^{rs}_{n0} e^{n \xi_r} \left( \hat b_n^{(r)} + \hat b_{-n}^{(r)} \right) \hat c^{(s)}_0 
+ \frac{1}{\sqrt{2}} \sum_{m\ge 1} \bar G^{rs}_{0m} e^{m \xi_s} \hat b^{(r)}_0 \left( \hat c_m^{(s)} + \hat c_{-m}^{(s)} \right) \nn\\
&& + \frac{1}{4} \bar G^{rs}_{(-1)(-1)} e^{\xi_r +\xi_s} \left( \hat b_1^{(r)} + \hat b_{-1}^{(r)} \right)  \left( \hat c_1^{(s)} + \hat c_{-1}^{(s)} \right) 
\nn\\
&& + \frac{1}{2} \sum_{m \ge 1} \bar G^{rs}_{(-1)m} e^{\xi_r +m\xi_s} \left( \hat b_1^{(r)} + \hat b_{-1}^{(r)} \right)  \left( \hat c_m^{(s)} + \hat c_{-m}^{(s)} \right) \nn\\
&&+ \frac{1}{2} \sum_{n \ge 1} \bar G^{rs}_{n(-1)} e^{n\xi_r +\xi_s} \left( \hat b_n^{(r)} + \hat b_{-n}^{(r)} \right)  \left( \hat c_1^{(s)} + \hat c_{-1}^{(s)} \right) \nn\\
&& + \sum_{n,m \ge 1} \bar G^{rs}_{nm} e^{n\xi_r +m\xi_s} \left( \hat b_n^{(r)} + \hat b_{-n}^{(r)} \right)  \left( \hat c_m^{(s)} + \hat c_{-m}^{(s)} \right)
\Biggr\}.
\eeq 
Here, $\hat b^{(r)}_n$ and $\hat c^{(r)}_m$ are anticommuting operators satisfying
\beq
\{\hat b^{(r)}_n, \hat c^{(r)}_m\} = \d^{rs} \d(n+m), ~~ r, s = 1, 2, 3. 
\eeq 

Finally, we are in a position to define the vertex operator in the ghost sector $\vert V[3]_{\rm gh} \rangle$. Extracting the contribution of the free propagation of the open strings from the string path integral, we obtain
\beq \label{vertex24}
\vert V[3]_{\rm gh} \rangle &=& \exp\left(-\sum_{r=1}^3 \xi_r L^{(r)}_{0,{\rm gh}} \right)
\exp \left(F^\prime_{\rm ghost}  \right) \vert 0 \rangle\nn\\
&=&  \sum_{r,s=1}^3 \Biggl\{\frac{1}{2}\bar G^{rs}_{00} \hat b^{(r)}_0 \hat c^{(s)}_0 + \frac{1}{2\sqrt{2}} \bar G^{rs}_{(-1)0} \hat b_{-1}^{(r)} \hat c^{(s)}_0 + \frac{1}{2\sqrt{2}} \bar G^{rs}_{0(-1)} \hat b^{(r)}_0 \hat c_{-1}^{(r)}  \nn\\
&&+ \frac{1}{\sqrt{2}} \sum_{n\ge 1} \bar G^{rs}_{n0} \hat b_{-n}^{(r)} \hat c^{(s)}_0 
+ \frac{1}{\sqrt{2}} \sum_{m\ge 1} \bar G^{rs}_{0m}  \hat b^{(r)}_0 \hat c_{-m}^{(s)} + \frac{1}{4} \bar G^{rs}_{(-1)(-1)}  \hat b_{-1}^{(r)} \hat c_{-1}^{(s)} 
\nn\\
&& + \frac{1}{2} \sum_{m \ge 1} \bar G^{rs}_{(-1)m} \hat b_{-1}^{(r)} \hat c_{-m}^{(s)} + \frac{1}{2} \sum_{n \ge 1} \bar G^{rs}_{n(-1)} \hat b_{-n}^{(r)}   \hat c_{-1}^{(s)}  \nn\\
&& +  \sum_{n,m \ge 1} \bar G^{rs}_{nm}  \hat b_{-n}^{(r)}  \hat c_{-m}^{(s)} 
\Biggr\} \vert 0 \rangle.
\eeq 

\section{Faddeev--Popov Ghost of non-Ableian Gauge Field Theory and BRST Ghosts in Open String Theory}

We expand the string state in the ghost sector to identify the Faddeev--Popov ghost in the asymptotic region (in cylindrical space) as follows:
\beq \label{expansion}
\vert \Psi_{\rm gh} \rangle  &=&  \left\{ \bar\eta_0 (x) b_{0} + \bar\eta_1(x) b_{-1} + \bar\eta_{2}(x) b_{-2} + \cdots + \chi_0(x) c_0+ \chi_1(x) c_{-1} + \cdots    \right\} \vert 0 \rangle. 
\eeq 
Note that the component fields $\bar\eta_1(x)$ and $\chi_1(x)$ are massless. From the kinetic 
term of the string field $\langle \Psi_{\rm gh} \vert Q \vert \Psi_{\rm gh} \rangle$,
the component fields obtain their kinetic terms $\chi_1 \p \bar\eta_1 + \bar \eta_1 \p \chi_1$. 

If we collect the massless component field terms in the vertex, Eq. (\ref{vertex24}), 
we obtain 
\beq \label{ghost29}
\left\{\frac{1}{4}\bar G^{rs}_{(-1)(-1)} + \frac{1}{2} \bar G^{rs}_{(-1)1} +
\frac{1}{2} \bar G^{rs}_{1(-1)} + \bar G^{rs}_{11}
\right\}  b_{-1}^{(r)} c_{-1}^{(s)} \vert 0 \rangle. 
\eeq

As we demonstrate in the Appendix, 
\beq
\bar G^{rs}_{(-1)(-1)} = \bar G^{rs}_{(-1)1} =\bar G^{rs}_{1(-1)} =\bar G^{rr}_{11} =0. 
\eeq 
Thus, the BRST ghost field terms in Eq. (\ref{ghost29}) reduce to the following term:
\beq
\bar G^{rs}_{11}  b_{-1}^{(r)} c_{-1}^{(s)} \vert 0 \rangle, ~~\text{for}~~ r\not=s.
\eeq

To calculate the gauge--ghost field coupling, we 
choose the string state as follows:
\beq \label{external}
\langle \Psi^{(1)},  \Psi^{(2)},  \Psi^{(3)} \vert &=& \langle 0 \vert \prod_{r=1}^3
\left(\bolA(r)+ \bar \boleta(r) + \bar \bolchi(r)\right),
\eeq
where 
\beq
\bolA(r) = A_\m(p^{(r)}) a^{(r)\m}_1, ~~~ \bar \boleta (r) = \bar\eta(p^{(r)}) b^{(r)}_1, ~~~ \bolchi(r) = \chi(p^{(r)}) c^{(r)}_1.
\eeq 
Here $\bar\eta(p^{(r)})$ and $\chi(p^{(r)})$ are the massless component fields in Eq. (\ref{expansion}). To obtain symmetric cubic couplings, we need to symmetrize the external string states: If we choose the exteral string state in the gauge-ghost sector as in Eq. (\ref{external}), we will obtain the symmetrized combination as explicitly written in the following equation. 

Now the gauge--ghost field coupling terms can be written as 
\beq
S_{A\bar\eta \chi} &=&\int \prod_{r=1}^3 dp^{(r)} \d \left(\sum_{r=1}^3 p^{(r)} \right) {\rm tr} 
\langle 0 \vert \prod_{r=1}^3\left(\bolA(r)+ \bar \boleta(r) + \bolchi(r)\right) \nn\\
&&
\exp\left\{ \sum_{r,s =1}^3 \bar G^{rs}_{11} b^{(r)}_{-1} c^{(s)}_{-1} + \sum_{r,s=1}^3 \bar N^{rs}_{10} a^{(r)\dag}_1 \cdot p^{(s)} \right\} \vert 0 \rangle \nn\\
&=& \int \prod_{r=1}^3 dp^{(r)} \d \left(\sum_{r=1}^3 p^{(r)} \right) {\rm tr}  \Bigl(\bolA(1) \bar\boleta (2) \bolchi (3) + \bolA(1) \bolchi(2)\bar\boleta(3) \
\nn\\
&& + \bar \boleta(1) \bolA(2) \bolchi(3) + \bolchi(1) \bolA(2) \bar\boleta(3)+ \bar \boleta(1) \bolchi(2) \bolA(3)+ \bolchi(1) \bar\boleta(2) \bolA(3) \Bigr)
\nn\\
&& \left(\sum_{r,s=1}^3 \bar N^{rs}_{10} a^{(r)\dag}_1 \cdot
p^{(s)} \right) \left(\sum_{r,s =1}^3 \bar G^{rs}_{11} b^{(r)}_{-1} c^{(s)}_{-1} \right)
\vert 0 \rangle.
\eeq

It can be further rewritten as
\beq \label{SAeta1}																			
S_{A\bar\eta \chi} 
&=&
\int \prod_{r=1}^3 dp^{(r)} \d \left(\sum_{r=1}^3 p^{(r)} \right) {\rm tr}  \Bigl(\bolA(1) \bar\boleta (2) \bolchi (3) + \bolA(1) \bolchi(2)\bar\boleta(3) \nn
\\
&& -  \bolA(2) \bolchi(3) \bar \boleta(1)-  \bolA(2)\bar\boleta(3)\bolchi(1) +  \bolA(3)\bar \boleta(1) \bolchi(2)+ \bolA(3) \bolchi(1) \bar\boleta(2)\Bigr)
\nn\\
&& \left(\sum_{r,s=1}^3 \bar N^{rs}_{10} a^{(r)\dag}_1 \cdot
p^{(s)} \right) \left(\sum_{r,s =1}^3 \bar G^{rs}_{11} b^{(r)}_{-1} c^{(s)}_{-1} \right)
\vert 0 \rangle \nn\\
&=&\int \prod_{r=1}^3 dp^{(r)} \d \left(\sum_{r=1}^3 p^{(r)} \right) {\rm tr}  \Biggl\{ \sum_{s=1}^3 \bar N^{1s}_{10} A(1) \cdot p^{(s)} \bar G^{32}_{11} \bar \eta(2) \chi(3) \nn\\
&& + \sum_s \bar N^{1s}_{10} A(1) \cdot p^{(s)} \bar G^{23}_{11}  \chi(2) \bar \eta(3)
-\sum_s \bar N^{2s}_{10} A(2) \cdot p^{(s)} \bar G^{31}_{11} \chi(3) \bar \eta(1)
\nn\\&&
-\sum_s \bar N^{2s}_{10} A(2) \cdot p^{(s)} \bar G^{13}_{11} \bar \eta(3) \chi(1)
+\sum_{s=1}^3 \bar N^{3s}_{10} A(3) \cdot p^{(s)} \bar G^{21}_{11} \bar \eta(1) \chi(2)
\nn\\
&& +\sum_s \bar N^{3s}_{10} A(3) \cdot p^{(s)} \bar G^{12}_{11} \bar\chi(1) \eta(2)  \Biggr\}.
\eeq

Here, we make use of the previous results on open string Neumann functions \cite{Lee2022cubicm}
\beq \label{Lee2022cubic}
\bar N^{11}_{10} p^{(1)} + N^{12}_{10} p^{(2)} + N^{13}_{10} p^{(3)} &=& \frac{2}{3\sqrt{3}} \left(p^{(2)} - p^{(3)} \right), \nn\\
\bar N^{21}_{10} p^{(1)} + N^{22}_{10} p^{(2)} + N^{23}_{10} p^{(3)} &=& \frac{2}{3\sqrt{3}} \left(p^{(3)} - p^{(1)} \right), \nn\\
\bar N^{31}_{10} p^{(1)} + N^{32}_{10} p^{(2)} + N^{33}_{10} p^{(3)} &=& \frac{2}{3\sqrt{3}} \left(p^{(1)} - p^{(2)} \right).
\eeq 

From calculations of $\bar G^{rs}_{11}$ given in the Appendix, 
we have 
\beq\label{G1234}
\bar G^{12}_{11} &=& \frac{2^8}{3^3 \sqrt{3}}, ~~ \bar G^{13}_{11} = \frac{2^6}{3^3 \sqrt{3}},~~ \bar G^{21}_{00} = 0, \nn\\
\bar G^{23}_{11} &=& 0, ~~ \bar G^{31}_{11} = - \frac{2^6}{3^3 \sqrt{3}}, ~~ \bar G^{32}_{11} = - \frac{2^8}{3^3\sqrt{3}}. 
\eeq 

Combining Eq. (\ref{Lee2022cubic}) and Eq. (\ref{G1234}) into Eq. (\ref{SAeta1}), 
we obtain
\beq
S_{A\bar\eta \chi} &=&\int \prod_{r=1}^3 dp^{(r)} \d \left(\sum_{r=1}^3 p^{(r)} \right) \frac{2}{3\sqrt{3}} \frac{2^6}{3^3 \sqrt{3}}{\rm tr} \Bigl\{ \left(p^{(2)}- p^{(3)}\right)(-4) \cdot A(1) \bar \eta(2) \chi(3) \nn\\
&& -  \left(p^{(3)}- p^{(1)}\right)(-1) \cdot A(2) \chi(3) \bar \eta(1) -  \left(p^{(3)}- p^{(1)}\right) \cdot A(2) \bar \eta(3) \chi (1) \nn\\
&& +\left(p^{(1)}- p^{(2)}\right) 4 \cdot A(3) \chi(1) \bar\eta(2) \Bigr\} \nn\\
&=& \int \prod_{r=1}^3 dp^{(r)} \d \left(\sum_{r=1}^3 p^{(r)} \right) \left(\frac{2^7}{3^5}\right) \Bigl\{ -4 \left(p^{(2)}- p^{(3)}\right)\cdot A(1) \bar\eta(2) \chi(3) \nn\\
&& \left(p^{(3)}- p^{(1)}\right) \cdot A(2) \chi(3) \bar \eta(1) -\left(p^{(3)}- p^{(1)}\right) \cdot A(2)  \bar\eta(3) \chi(1) \nn\\
&&
+ 4 \left(p^{(1)}- p^{(2)}\right) \cdot A(3) \chi(1) \bar \eta(2)  \Bigr\} \nn\\
&=& \int \prod_{r=1}^3 dp^{(r)} \d \left(\sum_{r=1}^3 p^{(r)} \right) 
\left(\frac{2^7}{3^5}\right)
\Bigl\{ 4 \left(p^{(3)}- p^{(2)}\right)\cdot A(1) \bar\eta(2) \chi(3) \nn\\
&& \left(p^{(3)}- p^{(2)}\right) \cdot A(1) \chi(3) \bar \eta(2) +\left(p^{(3)}- p^{(2)}\right) \cdot A(1)  \bar\eta(2) \chi(3) \nn\\
&&
+ 4 \left(p^{(3)}- p^{(2)}\right) \cdot A(1) \chi(3) \bar \eta(2)  \Bigr\} \nn\\
&=& \int \prod_{r=1}^3 dp^{(r)} \d \left(\sum_{r=1}^3 p^{(r)} \right) 
\left(\frac{2^7}{3^5}\right)
5 \left(p^{(3)}- p^{(2)}\right) \cdot \Bigl\{  A(1) \bar\eta(2) \chi(3) + A(1) \chi(3) \bar \eta(2)\Bigr\} \nn\\
&=& -g \int d^d x f_{abc} \p_\m \bar \eta^a A_\m^b \chi^c, 
\eeq 
where $g = \left(\frac{2^6}{3^5}\right)5$.
This is the cubic interction part of the Faddeev--Popov ghost action 
\beq
S_{\rm FP} &=& \int d^d x \p_\m \bar \eta^a \left(D^\m \chi \right)^a = \int d^d x \p_\m \bar
\eta^a \left(\p^\m \d^{ac} -g f_{abc} A^{\m b} \right) \chi^c.
\eeq

\section{Conclusions and Discussions}

In this study, we investigated the BRST ghost coordinate fields for open strings on multiple $Dp$-branes. On multiple $Dp$-branes, the string fields carry non-Abelian group indices and the massless components become non-Abelian gauge fields. In string field theory, the local gauge symmetry appears to be fixed covariantly; hence, the role of the ghost field, the massless component of the BRST ghost coordinates, becomes important to maintain the local gauge invariance. Siegel asserted that the massless component of the BRST ghost coordinates corresponds to the Faddeev--Popov ghost in gauge field theory. However, his assertion was tested only
for $U(1)$ gauge theory, which is the local gauge symmetry on a single $D25$-brane. It becomes a nontrivial task to verify multiple $Dp$-branes, where
the local gauge symmetry is a non-Abelian $SU(N)$ gauge symmetry and the Faddeev--Popov ghost field couples nontrivially with the non-Abelian gauge field. 

For this purpose, we constructed a ghost vertex operator for the open string on multiple $Dp$-branes and evaluated the coupling of the massless components of the BRST ghost coordinates and non-Abelian gauge field. The first step was to
construct Green's function (propagator) for the $b$-$c$ ghost coordinates on the string world sheet to evaluate the three-string scattering amplitude using the Polyakov string path integral. Because we had already established that Green's functions for the
$b$-$c$ ghosts (holomorphic and anti-holomorphic parts separately) of the closed string assume a simple form, we were only required to determine how the $b$-$c$ ghost coordinates transform under the conformal transformation; their propagator is written in terms of the closed string $b$-$c$ ghost coordinates. The answer was found in a previous work where an open string could be described as a folded closed string \cite{Lee1988Ann} and a recent work \cite{Lee2022cubicm,Lee2022cubic}, where an explicit form of the Schwarz--Christoffel mapping, from the string world sheet to a complex plane, was given. The massless
component fields of the BRST ghost are indeed the Faddeev--Popov ghost of the non-Abelian
gauge field, corresponing to the massless component of the open string field on multiple
$Dp$-branes.

We must emphasize that the vertex operator was not obtained by directly rewriting the overlapping function in terms of the oscillatory basis, as mentioned earlier.
This procedure can lead to different results; for example, on multiple $Dp$-branes, it does not produce the correct cubic gauge coupling. The reason is 
that a string does not propagate freely when it leaves the overlapping region. The spatial 
boundary condition set by the overlapping condition continues to influence string propagation. The string is not a point object. The correct procedure involved evaluating the string scattering amplitude and rewriting it in terms of the operators. We then extracted the contribution of the free
string propagation yielding the vertex operator. 

This paper creates more answers than it answers: We employ Witten's open string theory 
to study the $b$-$c$ ghost sector in string field theory. We confirmed that the massless
component field plays the role of the Faddeev--Popov ghost of the non-Abelian gauge theory, such that local non-Abelian gauge symmetry is maintained in the low-energy region. However, the coupling constants of the ghost gauge fields and those of the cubic gauge fields
do not agree. This could be due to the conic singularity of Witten's string field theory. It has also been noted before that the coupling of cubic gauge fields and that of quartic gauge fields are not in agreement with each other \cite{Moeller2000,Taylor2000}; this was in a study of Witten's cubic string field theory using the level truncation method. This motivates us to study the ghost sector of cubic 
string field theory in a proper-time gauge \cite{TLeeJKPS2017,Lee2017d,TLee2017cov,Lai2018,TLeeEPJ2018,Lee2019PLBfour,Lee2019four}. In a proper-time gauge, the coupling of cubic gauge fields and that of quartic gauge fields agree with each other \cite{Lee2019PLBfour}.

The second issue this paper brings is the extension of this work to closed string field theory. It would be interesting to determine if the Kawai--Lewellen--Tye (KLT) relation \cite{Kawai1986} holds for the ghost sector, such that the relationship between the general and gauge covariances could be understood at a deeper level. This extension would complete our recent work on closed cubic string field theory \cite{Lee2022cubic}. 

I implicitly assume the Siegel gauge, which fixes the BRST invariance in a way compitable with the Lorentz gauge for gauge field.  It may be interesting to examine the effect of choosing different gauge such as the Schnabl gauge \cite{Schnabl2006,Fuji2007}, which simplifies the star product drastically. 

This work could also shed light on the double copy theory \cite{BernPRL2010,Oxburgh2013,Monteiro2014,Kim2020}, which is based on the proposal ``gravity = gauge $\times$ gauge". A classical solution to Einstein gravity could possibly be obtained as a product of two copies of the non-Abelian gauge theory. In this approach, 
the ghost sector could help us understand the relationship between the general covariance of gravity theory and local gauge invariance of non-Abelian gauge theory. These issues will be discussed in subsequent papers.

\vskip 1cm

\begin{acknowledgments}
	
	This work was supported by a National Research Foundation of Korea (NRF) grant
	funded by the Korean government (MSIT) (2021R1F1A106299311). 
\end{acknowledgments}

\begin{appendix}
	
	\section{Neumann Functions of BRST Ghost Fields}
	
	On the string world sheet, the Neumann Green's function of the BRST can be defined as
	\beq \label{Grbrst200}
	G(\r_r, \rp_s) &=& \langle b(\r_r) c(\rp_s) \rangle_{\text{world sheet}} \nn, \\
	&=&
	\frac{1}{4} \left\{\left(\frac{\p z_r}{\p \zeta_r}\right)^{2} \left(\frac{\p \zp_s}{\p \zeta_s^\prime}\right)^{-1} 
	\frac{1}{z_r-\zp_s} + \left(\frac{\p z_r}{\p \zeta_r}\right)^{2} \left(\frac{\p \barzp_s}{\p \bar \zeta^\prime_s}\right)^{-1} \frac{1}{z_r-\barzp_s} + C.C. \right\},
	\eeq 
	where $\z_r$ are the local coordinates on the string world sheet of the $t$-th local patch, and $z_r$ are the local coordinates on the upper half of the complex plane. Based on Eq. (\ref{Grbrst200}), we expect that $G(\r_r, \rp_s)$ can be expanded in terms of local coordinates as
	\beq \label{Grbrst201}
	G(\r_r, \rp_s) &=& \d_{rs} \frac{1}{4} \sum_{n\ge 0} \left(\o_+^{-n+1} + (\bar\o_+)^{-n+1}
	\right) \left(\op_{}-^{n-1} + (\bar\o^\prime_-)^{n-1} \right) \nn\\
	&& + \frac{1}{4} \sum_{n, m \ge 0} \bar G^{rs}_{nm}  \left(\o_r^{n+2} + (\bar\o_r)^{n+2}
	\right) \left(\op_{}s^{m-1} + (\bar\o^\prime_s)^{m-1} \right), 
	\eeq 
	where $\o_r = e^{\zeta_r} = e^{\xi_r+ i\eta_r}, ~~~\op_s = e^{\xi^\prime_s + i\eta^{\prime}_s}$ and
	\beq
	(\o_+,\o_-) &=&  \left\{ 
	\begin{array}{ll}
		(\o_r, \op_s), & ~~\mbox{for} ~~\xi_r \ge \xi^\prime_s   \\ 
		(\op_s, \o_r), & ~~\mbox{for} ~~\xi_r \le \xi^\prime_s ~~.
	\end{array} \right. ~~~
	\eeq
	
	For Witten's BRST ghost field, the conformal mapping from the string world sheet, described by 
	$\zeta_r$, $r=1,2,3$, onto the upper half of the complex plane, denoted by $z_r$, is defined by the following two 
	consecutive mappings: \begin{subequations}
		\beq
		\omega_1 &=& e^{-\frac{2\pi i}{3}} \left(\frac{1+ i e^{\zeta_1}}{1 - i e^{\zeta_1}}\right)^{\frac{2}{3}},\\
		\omega_2 &=& \left(\frac{1+ i e^{\zeta_2}}{1 - i e^{\zeta_2}}\right)^{\frac{2}{3}}, \\
		\omega_3 &=& e^{\frac{2\pi i}{3}} \left(\frac{1+ i e^{\zeta_3}}{1 - i e^{\zeta_3}}\right)^{\frac{2}{3}}, 
		\eeq 
	\end{subequations}
	where the local coordinates of the three patches are given as $\zeta_r = \xi_r + i \eta_r$, $r= 1, 2, 3$.
	At the interaction point, $B$ is mapped to the origin of the disk and the external strings are located at 
	$e^{-\frac{2\pi i}{3}}, ~ 1, and ~ e^{\frac{2\pi i}{3}}$, respectively. In a compact form,
	\beq
	\omega_n &=& e^{\frac{2(n-2)}{3}\pi i} \left(\frac{1+ i e^{\zeta_n}}{1 - i e^{\zeta_n}}\right)^{\frac{2}{3}}, ~~~ n = 1, 2, 3.
	\eeq

	Then, each local coordinate patch on the unit disk is mapped onto the 
	upper half plane by the following conformal transformation:
	\beq
	z = -i \,\frac{\omega_r -1}{\omega_r +1}, ~~~
	\frac{\pi}{3} \le \arg\, \omega_r \le \frac{2\pi}{3}, ~~~ r =1, ~2,~ 3. 
	\eeq 
	The external strings are mapped to three points on the real line: $Z_n =\tan\left(\frac{n-2}{3} \pi \right)$, $n=1,2,3$, or explicitly,
	\beq
	Z_1 = -\sqrt{3}, ~~ Z_2 = 0, ~~~ Z_3 = \sqrt{3}.
	\eeq 
	The Schwarz–-Christoffel mapping from the local coordinate patch on the string work sheet to the upper (lower) half complex plane is expressed by series expansions \cite{Lee2022cubic} 
	\beq\label{expand46}
	e^{-\z_r} &=& \frac{a_r}{(z_r-Z_r)} + \sum_{n=0} c^{(r)}_n (z_r-Z_r)^n \nn\\
	a_1 &=& \frac{8}{3}, ~~ a_2 = \frac{2}{3}, ~~~ a_3 = \frac{8}{3}, \nn\\
	c^{(1)}_0 &=&\frac{2\sqrt{3}}{3}, ~~ c^{(1)}_1 = - \frac{5}{72}, ~~c^{(1)}_2 = \frac{5\sqrt{3}}{288} \nn\\
	c^{(2)}_0 &=& 0, ~~ c^{(2)}_1 = - \frac{5}{18}, ~~ c^{(2)}_2 = 0,\nn\\
	c^{(3)}_0 &=& -\frac{2\sqrt{3}}{3}, ~~ c^{(3)}_1 = - \frac{5}{72}, ~~c^{(3)}_2 = -\frac{5\sqrt{3}}{288}. \nn 
	\eeq

	\section{Calculation of Neumann Functions}
	
	\begin{itemize} 
		
		\item $\bar G^{rs}_{00}$ for $r \not=s$: We take
		a limit of Eq. (\ref{Grbrst200}) 
		where $z \rightarrow Z_r$ and $\zp \rightarrow Z_s$:
		\beq \label{br235}
		G(\r_r, \rp_s) &\rightarrow& \frac{1}{4} \left\{ \frac{(z_r - Z_r)^2}{(\zp_s-Z_s)} \frac{1}{Z_r-Z_s} + \frac{(z_r - Z_r)^2}{(\barzp_s-Z_s)} \frac{1}{Z_r-Z_s} + C. C. \right\} \nn\\
		&=& \frac{1}{4(Z_r-Z_s)} \left(\frac{\o^2_r}{\op_s} + \frac{\o^2_r}{\bar\o^\prime_s}
		+ C. C. \right).
		\eeq 
		Here, we note that as $z_r \rightarrow Z_r$ and $z^\prime_s \rightarrow Z_s$, 
		\beq
		\o_r = e^{\z_r} \rightarrow z_r -Z_r, ~~~ \o^\prime_s= e^{\z^\prime_s} \rightarrow z^\prime_s -Z_s.
		\eeq 
		Thus, a comparison of Eq. (\ref{br235}) with Eq. (\ref{Grbrst201}) yields
		\beq
		\bar G^{rs}_{00} = \frac{1}{Z_r-Z_s} = \frac{1}{\tan\left(\frac{r-2}{3} \pi \right) - \tan\left(\frac{s-2}{3} \pi \right)}. 
		\eeq 
		To be explicit, 
		\beq \label{g00238}
		\bar G^{12}_{00} = - \frac{\sqrt{3}}{3}, ~~\bar G^{13}_{00} = - \frac{\sqrt{3}}{6}, ~~ \bar G^{21}_{00} =   \frac{\sqrt{3}}{3}, ~~\bar G^{23}_{00} = - \frac{\sqrt{3}}{3}, ~~ \bar G^{31}_{00} =  \frac{\sqrt{3}}{6}, ~~\bar G^{32}_{00} = \frac{\sqrt{3}}{3}.
		\eeq
		
		\item $\bar G^{rs}_{n0}$, $n \ge 0$: Differentiating Eq. (\ref{Grbrst200}) and Eq. (\ref{Grbrst201}) with respect to $\zeta_r$,
		\beq \label{frs239}
		\frac{\p}{\p \zeta_r} G(\r_r, \rp_s) &=& \frac{1}{4} \left(\frac{\p z_r}{\p \zeta_r}\right) \frac{\p}{\p z_r}
		\left\{\left(\frac{\p z_r}{\p \zeta_r}\right)^{2} \left(\frac{\p \zp_s}{\p \zeta_s^\prime}\right)^{-1} 
		\frac{1}{z_r-\zp_s} + \left(\frac{\p z_r}{\p \zeta_r}\right)^{2} \left(\frac{\p \barzp_s}{\p \bar \zeta^\prime_s}\right)^{-1} \frac{1}{z_r-\barzp_s} \right\}
		\nn\\
		&=& \d_{rs} \frac{1}{4} \sum_{n\ge0} (-n+1) \o_r^{-n+1} \left(\op_s{}^{n-1} + (\bar\o^\prime_s)^{n-1} \right)  \nn\\
		&&~ + \frac{1}{4} \sum_{n, m\ge 0} \bar G^{rs}_{nm} (n+2) \o^{n+2}_r \left(\op_s{}^{m-1} + (\bar\o^\prime_s)^{m-1} \right),
		\eeq 
		Taking the limit where $\zp_s \rightarrow Z_s$ ($\op_s \rightarrow 0$) (the leading term is proportional to $1/\o^\prime_s$), 
		\beq\label{frs240}
		\left(\frac{\p z_r}{\p \zeta_r}\right)\frac{\p }{\p z_r} \left\{\left(\frac{\p z_r}{\p \zeta_r}\right)^2 \frac{1}{z_r -Z_s}
		\right\} = \d_{rs} \o_r+ \sum_{n\ge 0} \bar G^{rs}_{n0} (n+2) \o^{n+2}_r. 
		\eeq 
		Performing a contour integral around $\o_r = 0$ ($z_r= Z_r$), for $n \ge 0$,
		\beq
		&&\oint_{\o_r=0} d\o_r \o^{-n-3}_r \left\{\d_{rs} \o_r+ \sum_{n\ge 0} \bar G^{rs}_{n0} (n+2) \o^{n+2}_r\right\} \nn\\
		&& = \oint_{\o_r=0} d\o_r \o^{-n-3}_r \left(\frac{\p z_r}{\p \zeta_r}\right)\frac{\p }{\p z_r} \left\{\left(\frac{\p z_r}{\p \zeta_r}\right)^2 \frac{1}{z_r -Z_s}
		\right\}.
		\eeq 
		\beq
		\bar G^{rs}_{n0} = \frac{1}{(n+2)} \oint_{Z_r} \frac{dz_r}{2\pi i} \frac{\p }{\p z_r}\left\{\left(\frac{\p z_r}{\p \zeta_r}\right)^2 \frac{1}{z_r -Z_s}
		\right\} \,e^{-(n+2)\zeta_r(z_r)}, ~~ n \ge 0.
		\eeq 
		If $r \not=s$, 
		\beq \label{frs243}
		\bar G^{rs}_{n0} = \frac{1}{(n+2)} \oint_{Z_r} \frac{dz_r}{2\pi i} \left\{ \frac{2(z_r-Z_r)}{Z_r-Z_s} - \frac{(z_r-Z_r)^2}{(Z_r-Z_s)^2}
		\right\}e^{-(n+2)\zeta_r(z_r)}.
		\eeq 
		If $r=s$, 
		\beq \label{frs244}
		\bar G^{rr}_{n0} = \frac{1}{(n+2)} \oint_{Z_r} \frac{dz_r}{2\pi i} \, e^{-(n+2)\zeta_r(z_r)}. 
		\eeq

		\item $\bar G^{rs}_{nm}$, $n \ge 0,  m  \ge 2$: We differentiate Eq. (\ref{frs239}) with respect to $\zeta^\prime_s$ to obtain $\bar G^{rs}_{nm}$,
		\beq \label{frs36}
		\frac{\p}{\p \zeta_r}\frac{\p}{\p \zeta^\prime_s}  G(\r_r, \rp_s) &=&\frac{1}{4} \left(\frac{\p z_r}{\p \zeta_r}\right) 
		\left(\frac{\p z^\prime_s}{\p \zeta^\prime_s}\right) 
		\frac{\p}{\p z_r}\frac{\p}{\p z^\prime_s}
		\left\{\left(\frac{\p z_r}{\p \zeta_r}\right)^{2} \left(\frac{\p \zp_s}{\p \zeta_s^\prime}\right)^{-1} \frac{1}{z_r-\zp_s}  \right\}\nn\\
		&=& -\d_{rs} \frac{1}{4} \sum_{n\ge0} (n^2-1) \o_r^{-n+1} \op_s{}^{n-1} \nn\\
		&&+ \frac{1}{4} \sum_{n, m \ge 0}(n+2)(m-1) \bar G^{rs}_{nm}  \o^{n+2}_r \op_s{}^{m-1}.
		\eeq
		Performing a contour integral $\oint d\o_r \oint d \o^\prime_s (\o_r)^{-n-3} (\op_s)^{-m}$
		around $\o_r = 0$ ($z_r= Z_r$) and $\op_s = 0$ ($z^\prime_s= Z_s$), we obtain 
		\beq
		\bar G^{rs}_{nm} &=& \frac{1}{(n+2)(m-1)} \oint_{Z_r} \frac{dz_r}{2\pi i} \oint_{Z_s} \frac{dz^\prime_s}{2\pi i} \frac{\p}{\p z_r}\frac{\p}{\p z^\prime_s}
		\left\{\left(\frac{\p z_r}{\p \zeta_r}\right)^{2} \left(\frac{\p \zp_s}{\p \zeta_s^\prime}\right)^{-1} \frac{1}{z_r-\zp_s}  \right\}\nn\\
		&&~~~~
		e^{-(n+2)\z_r(z_r)} e^{-(m-1)\zeta^\prime_s(z^\prime_s)}.~~~~~~~~
		\eeq 
		Note that this equation does not determine when $m=1$, $\bar G^{rs}_{n1}$.
		
		\item $\bar G^{rs}_{n 1}, ~~ n \ge 0.$
		
		We can rewrite Eq. (\ref{frs239}) as follows: 
		\beq 
		\frac{\p}{\p \zeta_r} G(\r_r, \rp_s) &=& \frac{1}{4} \left(\frac{\p z_r}{\p \zeta_r}\right) \frac{\p}{\p z_r}
		\left\{\left(\frac{\p z_r}{\p \zeta_r}\right)^{2} \left(\frac{\p \zp_s}{\p \zeta_s^\prime}\right)^{-1} 
		\frac{1}{z_r-\zp_s} + \left(\frac{\p z_r}{\p \zeta_r}\right)^{2} \left(\frac{\p \barzp_s}{\p \bar \zeta^\prime_s}\right)^{-1} \frac{1}{z_r-\barzp_s} \right\}
		\nn\\
		&=& \d_{rs} \frac{1}{4} \sum_{n\ge0} (-n+1) \o_r^{-n+1} \left(\op_s{}^{n-1} + (\bar\o^\prime_s)^{n-1} \right)  \nn\\		&&~ + \frac{1}{4} \sum_{n \ge 0, m\ge 1} \bar G^{rs}_{nm} (n+2) \o^{n+2}_r \left(\op_{}s^{m-1} + (\bar\o^\prime_s)^{m-1} \right), \nn \\
		&& ~ + \frac{1}{4} \sum_{n \ge 0} \bar G^{rs}_{n0} (n+2) \o^{n+2}_r \left(\op{}_s^{-1} + (\bar\o^\prime_s)^{-1} \right).
		\eeq 
		Equivalently,
		\beq \label{B14}
		&& \sum_{n \ge 0, m\ge 1} \bar G^{rs}_{nm} (n+2) \o^{n+2}_r \left(\op_{}s^{m-1} + (\bar\o^\prime_s)^{m-1} \right), \nn\\
		&=&\left(\frac{\p z_r}{\p \zeta_r}\right) \frac{\p}{\p z_r}
		\left\{\left(\frac{\p z_r}{\p \zeta_r}\right)^{2} \left(\frac{\p \zp_s}{\p \zeta_s^\prime}\right)^{-1} 
		\frac{1}{z_r-\zp_s} + \left(\frac{\p z_r}{\p \zeta_r}\right)^{2} \left(\frac{\p \barzp_s}{\p \bar \zeta^\prime_s}\right)^{-1} \frac{1}{z_r-\barzp_s} \right\}\nn\\
		&& + \d_{rs} \sum_{n\ge0} (n-1) \o_r^{-n+1} \left(\op{}_s^{n-1} + (\bar\o^\prime_s)^{n-1} \right) \nn\\
		&&
		-\sum_{n \ge 0} \bar G^{rs}_{n0} (n+2) \o^{n+2}_r \left(\op{}_s^{-1} + (\bar\o^\prime_s)^{-1} \right).
		\eeq 
		
		If we collect the holomorphic part of Eq. (\ref{B14}), then 
		\beq
		\sum_{n \ge 0, m\ge 1} \bar G^{rs}_{nm} (n+2) \o^{n+2}_r \op{}_s^{m-1} &=& \left(\frac{\p z_r}{\p \zeta_r}\right) \frac{\p}{\p z_r}
		\left\{\left(\frac{\p z_r}{\p \zeta_r}\right)^{2} \left(\frac{\p \zp_s}{\p \zeta_s^\prime}\right)^{-1} 
		\frac{1}{z_r-\zp_s} \right\} \nn\\
		&& +  \d_{rs} \sum_{n\ge0} (n-1) \o_r^{-n+1} \op{}_s^{n-1} \nn\\
		&& -\sum_{n \ge 0} \bar G^{rs}_{n0} (n+2) \o^{n+2}_r \op{}_s^{-1}.
		\eeq 
		Performing an integral $\oint d\o_r (\o_r)^{-n-3} \oint d\o^\prime_s (\o^\prime_s)^{-1} $ around $\o_r = 0$ ($z_r= Z_r$) and $\o^\prime_s =0$ ($z^\prime_s = Z_s$), 
		we obtain
		\beq
		\bar G^{rs}_{n1} &=& \frac{1}{(n+2)} \oint_{Z_r} \frac{dz_r}{2\pi i} \oint_{Z_s}\frac{dz^\prime_s}{2\pi i} 
		\left(\frac{\p \zp_s}{\p \zeta_s^\prime}\right)^{-1} \nn\\
		&& \frac{\p}{\p z_r}
		\left\{\left(\frac{\p z_r}{\p \zeta_r}\right)^{2} \left(\frac{\p \zp_s}{\p \zeta_s^\prime}\right)^{-1} 
		\frac{1}{z_r-\zp_s} \right\}  e^{-(n+2)\zeta_r(z_r)}.
		\eeq  
		In the limit $z_r \rightarrow Z_r$, $\frac{\p z_r}{\p \zeta_r} \rightarrow (z_r - Z_r)$, and 
		$z^\prime_s \rightarrow Z_s$, $\frac{\p z^\prime_s}{\p \zeta^\prime_s} \rightarrow (z^\prime_s - Z_s)$, we find 
		\beq
		\bar G^{rs}_{n1} &=& \frac{1}{(n+2)} \oint_{Z_r} \frac{dz_r}{2\pi i} \oint_{Z_s}\frac{dz^\prime_s}{2\pi i} 
		\frac{1}{(z^\prime_s - Z_s)} \frac{\p}{\p z_r}\left\{\frac{(z_r-Z_r)^2}{(z^\prime_s - Z_s)}
		\frac{1}{(z_r-\zp_s)} \right\}  e^{-(n+2)\zeta_r(z_r)} \nn\\
		&=& \frac{2}{(n+2)} \oint_{Z_r} \frac{dz_r}{2\pi i} \frac{(z_r-Z_r)}{ (z_r-Z_s)^3}(Z_r-Z_s) e^{-(n+2)\zeta_r(z_r)}.
		\eeq
		It vanishes for $r =s$,
		\beq
		\bar G^{rr}_{n1} = 0.
		\eeq 
		For $r\not =s$,
		\beq
		\bar G^{rr}_{n1} = \frac{2}{(n+2)} \oint_{Z_r} \frac{dz_r}{2\pi i} \frac{(z_r-Z_r)}{ (z_r-Z_s)^3}(Z_r-Z_s) e^{-(n+2)\zeta_r(z_r)}.
		\eeq

	\end{itemize}
	
	\section{Calculations of Neumann Functions, $\bar G^{rs}_{nm}$}
	
	\begin{itemize}
		
		\item $\bar G^{11}_{(-1)(-1)}=\bar G^{22}_{(-1)(-1)}=\bar G^{33}_{(-1)(-1)}$
		
		\beq 
		\bar G^{11}_{(-1)(-1)} &= - \frac{1}{2} &\oint_{Z_1} \frac{dz_1}{2\pi i} \oint_{Z_1} \frac{dz^\prime_1}{2\pi i} \frac{\p}{\p z_1}\frac{\p}{\p z^\prime_1}
		\left\{\left(\frac{\p z_1}{\p \zeta_1}\right)^{2} \left(\frac{\p \zp_1}{\p \zeta_1^\prime}\right)^{-1} \frac{1}{z_1-\zp_1}  \right\} e^{-\z_1(z_1)} e^{2\zeta^\prime_2(z^\prime_2)} \nn\\
		&=& \oint \frac{d z_1}{2\pi i} \oint \frac{d z^\prime_1}{2\pi i}\frac{\p}{\p z_1}\frac{\p}{\p z^\prime_1} \Biggl\{  \left( (z_1-\sqrt{3}) \right)^2\left( (z^\prime_1-\sqrt{3})  \right)^{-1}\frac{1}{z_1-\zp_1} \Biggr\} \nn\\
		&& \left(\frac{a_1}{(z_1-\sqrt{3}) } \right)  \left(\frac{a_1}{(z^\prime_1-\sqrt{3}) } \right)^{-2} \nn, \\
		&=& \oint_{Z_1} \frac{d z_1}{2\pi i} \oint_{Z_1} \frac{dz^\prime_1}{2\pi i} \frac{1}{a_1}\Biggl\{ - \frac{2}{(z_1-z^\prime_1)} + \frac{2(z^\prime_1 - \sqrt{3})}{(z_1-z^\prime_1)^2)} - \frac{(z_1-\sqrt{3})}{(z_1-z^\prime_1)^2}\nn\\
		&& - \frac{2(z_1-\sqrt{3})(z^\prime_1 - \sqrt{3})}{(z_1-z^\prime_1)^3}
		\Biggr\}=0.
		\eeq

		\item $\bar G^{rs}_{(-1)(-1)}=\bar G^{12}_{(-1)(-1)}$ for $r \not=s$ 
		
		\beq
		\bar G^{12}_{(-1)(-1)}
		&=&- \frac{1}{2}\oint \frac{d z_1}{2\pi i} \oint \frac{d z^\prime_2}{2\pi i}\frac{dz^\prime_1}{2\pi i} \frac{\p}{\p z_1}\frac{\p}{\p z^\prime_2}\left\{
		\left( (z_1-\sqrt{3})\right)^2
		\left(z^\prime_2  - \frac{5}{6}z^{\prime 3}_2 \right)^{-1}
		\frac{1}{z_1-\zp_2}  \right\}\nn\\
		&& \left(\frac{a_1}{(z_1-\sqrt{3}) } \right) \left(\frac{a_2}{z^\prime_2} \right)^{-2} \nn\\
		&=&\oint \frac{d z_1}{2\pi i}  \oint \frac{d z^\prime_2}{2\pi i}\frac{\p}{\p z_1}\frac{\p}{\p z^\prime_2}\left\{(z_1 - \sqrt{3})^2 \frac{1}{z^\prime_2} \frac{1}{z_1-z^\prime_2} \right\}\frac{a_1}{z_1-\sqrt{3}} \frac{z^{\prime 2}_2}{a^2_2} 
		= 0. 
		\eeq 
		
		\item $\bar G^{11}_{(-1)1}=\bar G^{22}_{(-1)1}=\bar G^{33}_{(-1)1} = 0.$

		\item $\bar G^{12}_{(-1)1}=\bar G^{23}_{(-1)1}=\bar G^{31}_{(-1)1}$
		
		\beq
		\bar G^{12}_{(-1)1}
		&=& 2\oint \frac{d z_1}{2\pi i} \frac{(z_1 - \sqrt{3})}{(z_1)^3} \sqrt{3} 
		\left(\frac{a_1}{(z_1-\sqrt{3}) } + \sum_{p=0} c^{(1)}_p (z_1-\sqrt{3})^p \right) \nn\\
		&=&2 \oint \frac{d z_1}{2\pi i} \left\{ \frac{(z_1 - \sqrt{3})^3}{a_1} \frac{a_2}{(z^\prime_2)^2} \frac{1}{z_1-\zp_2}\right\} =0,
		\eeq
		
		\item $\bar G^{21}_{(-1)1}=\bar G^{13}_{(-1)1}=\bar G^{32}_{(-1)1}$
		
		\beq
		\bar G^{21}_{(-1)1}
		&=- \frac{2}{\sqrt{3}}&\oint_{Z_2} \frac{d z_2}{2\pi i} \frac{z_2}{(z_2-\sqrt{3})^3}  \left(\frac{a_2}{z_2} + \sum_{q=0} c^{(2)}_q (z_2)^q \right) = 0,
		\eeq

		\item $\bar G^{11}_{1(-1)}=\bar G^{22}_{1(-1)}=\bar G^{33}_{1(-1)} = 0.$
		
		\beq
		\bar G^{11}_{1(-1)} &=&  - \frac{1}{6} \oint_{Z_1} \frac{dz_1}{2\pi i} \oint_{Z_1} \frac{dz^\prime_1}{2\pi i} \frac{\p}{\p z_1}\frac{\p}{\p z^\prime_1}
		\left\{\left(\frac{\p z_1}{\p \zeta_1}\right)^{2} \left(\frac{\p \zp_1}{\p \zeta_1^\prime}\right)^{-1} \frac{1}{z_1-\zp_1}  \right\} e^{-3\z_1(z_1)} e^{2\zeta^\prime_2(z^\prime_2)} \nn\\
		&=& \oint \frac{d z_1}{2\pi i} \oint \frac{d z^\prime_1}{2\pi i}\frac{\p}{\p z_1}\frac{\p}{\p z^\prime_1} \Biggl\{  \left( (z_1-\sqrt{3}) \right)^2\left( (z^\prime_1-\sqrt{3})  \right)^{-1}\frac{1}{z_1-\zp_1} \Biggr\} \nn\\
		&& \left(\frac{a_1}{(z_1-\sqrt{3}) } \right)^3  \left(\frac{a_1}{(z^\prime_1-\sqrt{3}) } \right)^{-2} \nn\\
		&=&  - \frac{1}{6} \oint_{Z_1} \frac{dz_1}{2\pi i} \oint_{Z_1}\frac{dz^\prime_1}{2\pi i} \Biggl\{
		-2 \frac{(z_1-\sqrt{3})}{(z^\prime-\sqrt{3})^2} \frac{1}{(z_1-z^\prime_1)} +\frac{(z_1-\sqrt{3})^2}{(z^\prime-\sqrt{3})^2} \frac{1}{(z_1-z^\prime_1)^2} \nn\\
		&&+ 2 \frac{(z_1-\sqrt{3})}{(z^\prime-\sqrt{3})} \frac{1}{(z_1-z^\prime_1)^2} -2 \frac{(z_1-\sqrt{3})^2}{(z^\prime-\sqrt{3})} \frac{1}{(z_1-z^\prime_1)^3}  \Biggr\}\nn \\
		&&\left(\frac{a_1}{(z_1-\sqrt{3}) } \right)^3  \left(\frac{a_1}{(z^\prime_1-\sqrt{3}) } \right)^{-2}\nn\\
		&=&0.
		\eeq
		
		\item $\bar G^{rs}_{1(-1)}=\bar G^{12}_{1(-1)}=0$ for $r \not=s$ 
		
		\beq
		\bar G^{12}_{1(-1)}
		&=&- \frac{1}{6}\oint \frac{d z_1}{2\pi i} \oint \frac{d z^\prime_2}{2\pi i}\frac{dz^\prime_1}{2\pi i} \frac{\p}{\p z_1}\frac{\p}{\p z^\prime_2}\left\{
		\left( (z_1-\sqrt{3}) \right)^2
		\left(z^\prime_2  \right)^{-1}
		\frac{1}{z_1-\zp_2}  \right\}\nn\\
		&& \left(\frac{a_1}{(z_1-\sqrt{3}) } \right)^3 \left(\frac{a_2}{z^\prime_2} \right)^{-2} \nn\\
		&=&- \frac{1}{6}\oint \frac{d z_1}{2\pi i} \oint \frac{d z^\prime_2}{2\pi i} \Biggl\{
		-2 \frac{(z_1-\sqrt{3})}{z^{\prime 2}_2(z_1-z^\prime_2)} + 2  \frac{(z_1-\sqrt{3})}{z^{\prime }_2(z_1-z^\prime_2)^2} \nn\\
		&& +  \frac{(z_1-\sqrt{3})^2}{z^{\prime 2}_2(z_1-z^\prime_2)^2} -2  \frac{(z_1-\sqrt{3})^2}{z^{\prime }_2(z_1-z^\prime_2)^3}
		\Biggr\}\left(\frac{a_1}{(z_1-\sqrt{3}) } \right)^3 \left(\frac{a_2}{z^\prime_2} \right)^{-2} 
		\nn\\
		&=& 0. 
		\eeq

		\item $\bar G^{11}_{11}=\bar G^{22}_{11}=\bar G^{33}_{11}=0.$
		
		\item $\bar G^{rs}_{11}$ for $r \not=s$:

		\noindent
		\beq
		\bar G^{12}_{11}
		&=&\frac{2}{3}\oint_{Z_1}\frac{d z_1}{2\pi i}
		\frac{(z_1-\sqrt{3})}{(z_1)^3} \sqrt{3}  \left(\frac{a_1}{(z_1-\sqrt{3}) } + \sum_{p=0} c^{(1)}_p (z_1-\sqrt{3})^p \right)^3 \nn\\
		&=& \frac{2}{\sqrt{3}} \frac{1}{(\sqrt{3})^3} 3 (a_1)^2 c^{(1)}_0 = \frac{2^8}{3^3 \sqrt{3}},
		\eeq 
		\beq
		\bar G^{23}_{11} &=& \frac{2}{3}\oint_{Z_2}\frac{d z_2}{2\pi i}
		\frac{z_2}{(z_2+\sqrt{3})^3} \sqrt{3}  \left(\frac{a_2}{z_2 } + \sum_{p=0} c^{(2)}_p (z_2)^p \right)^3  \nn\\
		&=& \frac{2}{3} \frac{\sqrt{3}}{(\sqrt{3})^3} 3 (a_2)^2 c^{(2)}_0 = 0.
		\eeq
		\beq
		\bar G^{31}_{11} &=& -\frac{2}{3}\oint_{Z_3}\frac{d z_3}{2\pi i}
		\frac{z_3+\sqrt{3}}{(z_3-\sqrt{3})^3} 2\sqrt{3}  \left(\frac{a_3}{z_3+ \sqrt{3} } + \sum_{p=0} c^{(3)}_p (z_3+ \sqrt{3})^p \right)^3  \nn\\
		&=& \frac{2^2 \sqrt{3}}{3} \frac{1}{(2\sqrt{3})^3} 3 (a_3)^2 c^{(3)}_0 = - \frac{2^6}{3^3\sqrt{3}}.
		\eeq
		\beq 
		\bar G^{21}_{11}
		&=&-\frac{2}{3}\oint_{Z_2}\frac{d z_2}{2\pi i}
		\frac{z_2}{(z_2-\sqrt{3})^3} \sqrt{3}  \left(\frac{a_2}{z_2 } + \sum_{p=0} c^{(2)}_p (z_2)^p \right)^3  \nn\\
		&=& - \frac{2}{\sqrt{3}} \frac{1}{(-\sqrt{3})^3} 3 (a_2)^2 c^{(2)}_0 = 0.
		\eeq
		\beq
		\bar G^{13}_{11}
		&=&\frac{2}{3}\oint_{Z_1}\frac{d z_1}{2\pi i}
		\frac{(z_1-\sqrt{3})}{(z_1+\sqrt{3})^3} 2\sqrt{3}  \left(\frac{a_1}{(z_1-\sqrt{3}) } + \sum_{p=0} c^{(1)}_p (z_1-\sqrt{3})^p \right)^3 \nn\\
		&=& \frac{2}{3} \frac{2\sqrt{3}}{(2\sqrt{3})^3} 3 (a_1)^2 c^{(1)}_0 = \frac{2^6}{3^3 \sqrt{3}},
		\eeq
		\beq
		\bar G^{32}_{11} &=& -\frac{2}{3}\oint_{Z_3}\frac{d z_3}{2\pi i}
		\frac{z_3+\sqrt{3}}{(z_2)^3} \sqrt{3}  \left(\frac{a_3}{z_3+ \sqrt{3} } + \sum_{p=0} c^{(3)}_p (z_3+ \sqrt{3})^p \right)^3  \nn\\
		&=& -\frac{2}{3} \frac{\sqrt{3}}{(-\sqrt{3})^3} 3 (a_3)^2 c^{(3)}_0 = - \frac{2^8}{3^3 \sqrt{3}}.
		\eeq

	\end{itemize}
	
\end{appendix}

\newpage


		%
	
		%


\end{document}